\documentclass[prl,aps,amssymb,twocolumn,showpacs,superscriptaddress]{revtex4}
\usepackage{graphicx}
\usepackage{amsmath}

\newcommand{\re}{\mathrm{Re}}
\newcommand\ket[1]{\left|{\textstyle#1}\right\rangle}

\begin{document}

\title{Decoherence Driven Quantum Transport}

\author{Sang Wook Kim}
\email{swkim0412@pusan.ac.kr}
\affiliation{Department of Physics
Education, Pusan National University, Busan 609-735, Korea}

\author{Mahn-Soo Choi}
\affiliation{Department of Physics, Korea University, Seoul 136-701, Korea}

\date{\today}

\begin{abstract}
We propose a new mechanism to generate a dc current of particles at zero
bias based on a noble interplay between coherence and decoherence. We
show that a dc current arises if the transport process in one direction
is maintained coherent while the process in the opposite direction is
incoherent.  We provide possible implementations of the idea using an
atomic Michelson interferometer and a ring interferometer.
\end{abstract}



\maketitle


The rates of emission and absorption
between two quantum states are equal as governed by the principle of
detailed balance.  Under special conditions, however,
the detailed balance can be broken,
and one of the processes can even be completely suppressed. One
relevant example is the so-called lasing without inversion (LWI) in
quantum optics \cite{Scully97a}. LWI is achieved in an ensemble of
atoms that have a pair of nearly degenerate ground-state levels,
say, $a$ and $b$ (Fig. 1). The atoms are prepared in a coherent
superposition of $a$ and $b$. This coherent superposition can be
realized by applying microwave resonant with the ground-state
splitting, $E_b-E_a$. The excitation probability to a upper level
$c$, $T_{ab \rightarrow c}$, undergoes two-path interference, and
hence can be modulated by changing the relative phase of the levels
$a$ and $b$ in the coherent superposition. On the contrary the
decaying probability from the state $c$ to both $a$ and $b$,
$T_{c \rightarrow ab}$, is just the summation of the probabilities
of two spontaneous decays since it is {\em incoherent} processes.
When the phase of the microwave field is adjusted to make $T_{ab
\rightarrow c}$ smaller than $T_{c \rightarrow ab}$, the lasing
operation is possible without population inversion.

\begin{figure}
\center
\includegraphics*{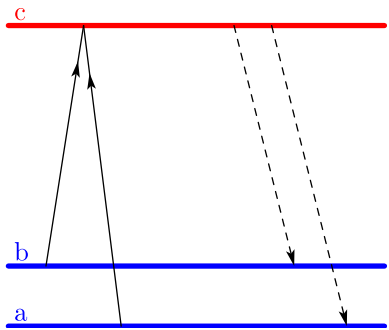}
\caption{(color online) The schematic diagram of the energy levels
of an atom for LWI
  operation. $a$, $b$ and $c$ represent the energy levels.}
\label{Paper::fig1}
\end{figure}

A \emph{dc} current of particles can be achieved by applying an
external \emph{dc} bias; e.g., an electrical potential difference
for charges, density difference for masses, temperature difference
for heat, and so on.  In these examples the external bias breaks
directly the detailed balance between the currents in opposite
directions.  It is also possible to obtain a dc current by breaking
the detailed balance \emph{indirectly}: In rectification, a
\textit{dc} current is generated by an external \emph{ac} voltage. In
``ratchets'', the directed motion of particles can be caused from
even random fluctuations \cite{Astumian02a}. Quantum mechanics
provides still another ways to generate dc current by applying ac
fields, e.g., (\emph{mesoscopic}) photo-voltaic effect
\cite{Falko89a}, quantum pumps \cite{Altshuler99a}, and
quantum version of classical ratchet \cite{Linke99a,Astumian02a}.

In this Letter we propose a new mechanism to generate dc currents at
zero-bias by generalizing the concept of LWI to the transport problem.
The basic idea is simple: A dc current will arise if the transport
process in one direction is coherent while the process in the
opposite direction is incoherent.  One can easily check the idea by
noting that the transmission probability of the coherent transport
varies with the relative phases of multiple paths, which does not
affect the incoherent one.
The important question is then how to realize such \emph{spatially}
anisotropic, coherent/incoherent transport processes.
It is clearly distinguished from the original idea of LWI since the
asymmetry in coherence/decoherence of our concern addresses spatial
directions rather than the excitation and relaxation of energy.

The scheme of our implementation is quite general, but for definiteness,
here we take two specific examples; one based on Michelson
interferometer and the other on a ring interferometer.

Let us first consider the scheme based upon the Michelson
interferometer; see Fig.~\ref{Paper::fig:Michelson-1,2,3}. We have
an atomic Michelson interferometer \cite{Dumke02a} and two
reservoirs, 1 and 2, of two-level atoms at the ends of the two
input/output channels of the interferometer.  The atoms from a
reservoir enter the interferometer, experience scattering and/or
interference, and is either reflected back to the original reservoir
or transmitted to the other. In addition we have an important
component, the microcavity (C) between the reservoir 2 and the
atomic beam splitter (BS)
\cite{Dumke02a}. The cavity is set
resonant to the level splitting $\Delta$ of the two-level atoms, so
that the atoms entering the cavity in the ground state come out of
the cavity in the excited state. Therefore, when entering the
interferometer, the atoms from the reservoir 2 are in the excited
state while those from the reservoir 1 remain in the ground state.
This difference in the energy state between the atoms entering the
interferometer can cause a significant difference in the coherence
of their center-of-mass (CM) motions in the interferometer. 

To see this, let $L_\tau=v\tau$, where $v$ is the velocity of the
atoms and $\tau$ is the lifetime of the excited energy level.
Provided that $L_\tau < 2L$, with $L$ being the lengths of the arms
(i.e., the paths from the atomic BS to the mirrors) of
the interferometer (the lengths of the arms are assumed to be equal), an
excited atom in the interferometer will relax
back to its ground state emitting a photon; see
Fig.~\ref{Paper::fig:Michelson-1,2,3}(b). In ideal case, the photon
enables us to locate the atom definitely on one of the two arms of
the interferometer.  The excited atoms thus never experience an
interference through the Michelson interferometer.  In this sense,
the CM motion of the atoms from reservoir 2 is \emph{incoherent}.
Furthermore, starting from the just located arm (whichever it is),
the atom is transmitted to reservoir 1 with probability 0.5 and
reflected back to reservoir 2 with 0.5 (we consider a 50:50
BS); see Fig.~\ref{Paper::fig:Michelson-1,2,3}(b). 

On the other hand, the atoms from reservoir 1 (ground-state atoms) do
not allow such relaxation and will experience coherent interference as
long as $L_\phi\gg2L$, where $L_\phi$ is the coherence length of the CM
motion of \emph{ground-state} atoms; see
Fig.~\ref{Paper::fig:Michelson-1,2,3} (a).  Due to the constructive
interference, an atom from reservoir 1 is perfectly transmitted to the
reservoir 2; see Fig.~\ref{Paper::fig:Michelson-1,2,3} (a).  Comparing
these two transport processes, one can see that 50\% of the incoming
atoms contribute to the net dc current. Namely, when the currents from
the reservoirs 1 and 2 are equal, $I_1=I_2=I$, the \emph{net} current
from 1 to 2 is given by
\begin{equation}
\label{Paper::eq:I:1}
I_{12} = I_1 - 0.5I_2 = 0.5I \,.
\end{equation}

\begin{figure}
\center
\includegraphics*[width=65mm]{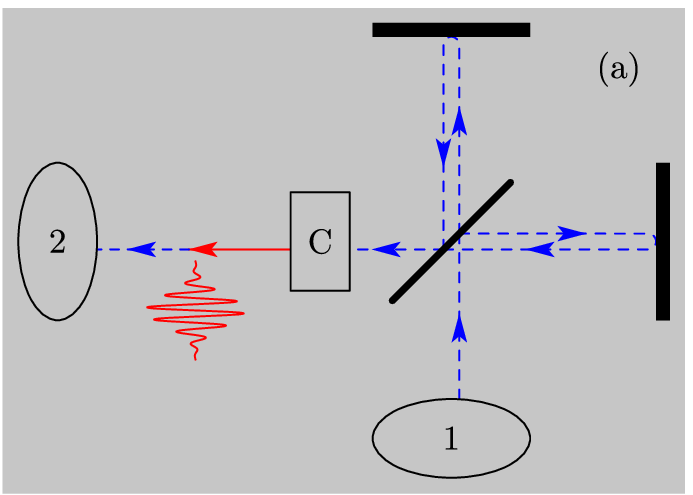}\\[2mm]
\includegraphics*[width=65mm]{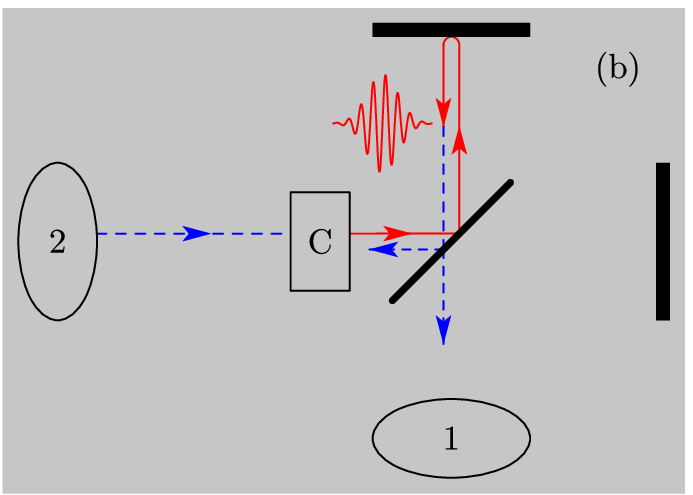}\\[2mm]
\caption{(color online) The scheme based on the Michelson
  interferometer. $1$ and $2$ represent reservoirs.  The (blue) dashed
  and the (red) solid lines represent the trajectories of the the
  ground-state and excited atoms, respectively.  The horizontal and the
  vertical thick lines are mirrors, and the titled thick lines in the
  middle an atomic beam splitter. The box with ``C'' is the
  microcavity. (a) The coherent process: An atom from the reservoir 1
  undergoes an constructive interference and reaches the reserves 2 with
  unit probability.  The cavity does not affect the transmission of this
  atom.  (b) The incoherent process: An atom from the reservoir 2 is
  excited at the cavity, and spontaneously emits a photon within the
  vertical path. The atom is then transmitted to either reservoir with
  equal probability 0.5.}
  \label{Paper::fig:Michelson-1,2,3}
\end{figure}

The current expressed in Eq.~(\ref{Paper::eq:I:1}) is the maximum
current attainable in our scheme assuming idealistic situations.
Certain imperfections in reality will diminish the current.
Firstly, only a fraction $P_{\mathrm{ex}}$ of the atoms from the
reservoir 2 may be excited by the cavity.
Secondly, an excited atom entering the interferometer may not
necessarily relax to the ground state \emph{inside} the interferometer.
The probability $P_\tau$ for such an event to occur is given by
\begin{math}
P_\tau \approx \int_0^{2L/v}{dt}\;e^{-t/\tau}
\end{math}
ignoring the distance between the BS and the cavity.
Thirdly, even if the excited atom relaxes inside the interferometer and
a photon is emitted, the atom cannot contribute to the net current
unless the photon gives enough information about which path of the
interferometer the atom takes.  For example, if the wavelength $\lambda$
of the photon is comparable to or larger than the size of the
interferometer $L$ ($\lambda\gtrsim L$), then one cannot get enough
which-path information and the CM motion still remains coherent.
Fourthly, while the decoherence due to photon emission is of particular
type and of our primary concern, in general, the CM motion is subject to
additional decoherence of usual type due to the ``environment'' even
when the atom keeps its internal state (either ground or excited).
Unlike the former, however, the latter is spatially isotropic and tends
to reduce the net current.

The simple inspections in Eq.~(\ref{Paper::eq:I:1}) and the effects of
the imperfections mentioned above can be treated more rigorously based
on the scattering theory\cite{Taylor72a}.
Important ingredients to be included in the formalism are the time
dependence and the wave-packet description. It is because the
decoherence process due to photon emission should locate the atom
within the interferometer and the subsequent scattering process of
the located atom is separated in time from that of the incoming
atom.  Another important element is an effective description of
decoherence.  To provide an unified description of both types of
decoherence (see above), we adopt the framework of the unitary
representation\cite{Nielsen00a}.
A 50:50 BS is described by the unitary scattering matrix (both for the
wave entering and leaving the interferometer):
\begin{equation}
\label{Paper::eq:S1}
S = \frac{1}{\sqrt{2}}
\begin{bmatrix}
1 & i \\
i & 1
\end{bmatrix} \,,
\end{equation}
ignoring the weak energy-dependence in the range of interest. The
state vector of the atom that has come from the reservoir 2
is given \emph{inside} the interferometer
by\cite{Taylor72a,Nielsen00a}
\begin{multline}
\label{Paper::eq:packet1}
\ket{\Psi(t)} = \int\frac{dk}{\sqrt{2\pi}}\;\phi(k)e^{-i\omega(k)t}
\times\\ \Bigg[
(\alpha\ket{g} + \beta e^{-i\Delta t/\hbar}\ket{e})\otimes\ket{0}\otimes
\frac{ie^{+ikx_1}\ket{e_1} + e^{+ikx_2}\ket{e_2}}{\sqrt{2}} \\\mbox{}\hfil%
+ \gamma\ket{g}\otimes
\frac{ie^{+ikx_1}\ket{\nu_1}\otimes\ket{e_1}
  + e^{+ikx_2}\ket{\nu_2}\otimes\ket{e_2}}{\sqrt{2}}
\Bigg] ,
\end{multline}
where $\phi(k)$ is the envelope function of the wave packet, $\omega(k)$
is the dispersion of the CM motion, $x_1$ ($x_2$) is the position from
the BS along the vertical (horizontal) arm of the interferometer.
$\ket{g}$ ($\ket{e}$) is the internal ground (excited) state of the
atom.  The photon emitted from the vertical (horizontal) arm is
represented by $\ket{\nu_1}$ ($\ket{\nu_2}$) whereas $\ket{0}$ is the
vacuum state. $\langle{\nu_1|\nu_2}\rangle=0$ implies that the photon
provides a sufficient which-path information and one can locate
perfectly the atom on one of the two arms.
The environment that couples to the CM motion and causes the
decoherence of usual type has the state $\ket{e_1}$ ($\ket{e_2}$)
when the atom takes the vertical (horizontal)
arm\cite{Stern90a,Moskalets01a}. $\langle{e_1|e_2}\rangle=0$ means
that the CM motion is completely incoherent even when the atom
keeps the same internal state. 
The coefficients $\alpha$, $\beta$, and $\gamma$ are related to the
probabilities $P_\mathrm{ex}$ and $P_\tau$ by
\begin{math}
1 - P_\mathrm{ex} = |\alpha|^2
\end{math},
\begin{math}
P_\mathrm{ex} (1 - P_\tau) = |\beta|^2
\end{math}, and
\begin{math}
P_\mathrm{ex} P_\tau = |\gamma|^2
\end{math}.
In Eq.~(\ref{Paper::eq:packet1}) we have assumed that the atom keeps the
same shape of the CM wave packet before and after the emission of a
photon.  This is valid when the packet size is already small compared
with $L$, and (to avoid the recoil of the atom when emitting a photon)
the CM momentum $\hbar{k}$ of the atom is sufficiently larger than that
of the photon $h/\lambda$.
The atom scatters again off the BS to get out of the interferometer and
then has the state vector
\begin{multline}
\label{Paper::eq:packet2}
\ket{\Psi(t)} = i\int\frac{dk}{\sqrt{2\pi}}\;
\phi(k)e^{i2kL-i\omega(k)t} \\ \times%
\Bigg[
\frac{e^{-ikx_1}}{2}
(\alpha\ket{g}+\beta e^{-i\Delta{t}/\hbar}\ket{e})
\otimes\ket{0}\otimes(\ket{e_1}+\ket{e_2}) \\\mbox{}%
+ \frac{ie^{-ikx_2}}{2}
(\alpha\ket{g}+\beta e^{-i\Delta{t}/\hbar}\ket{e})
\otimes\ket{0}\otimes(\ket{e_1}-\ket{e_2}) \\\mbox{}%
+ \frac{\gamma e^{-ikx_1}}{2}
\ket{g}\otimes(\ket{\nu_1}\otimes\ket{e_1}+\ket{\nu_2}\otimes\ket{e_2})
\\\mbox{}%
+ \frac{i\gamma e^{-ikx_2}}{2}
\ket{g}\otimes(\ket{\nu_1}\otimes\ket{e_2}-\ket{\nu_2}\otimes\ket{e_2})
\Bigg] .
\end{multline}
Therefore
the probability that an atom from the reservoir 2 reach the
reservoir 1 is given by
\begin{math}
P(1\leftarrow 2) =
(|\alpha|^2 + |\beta|^2)(1+\langle{e_1|e_2}\rangle)/2
+ |\gamma|^2(1+\re\langle{e_1|e_2}\rangle\langle{\nu_1|\nu_2}\rangle)/2
\end{math}, which leads to the net current
\begin{equation}
\label{Paper::eq:I:5}
I_{12}=0.5 P_{\mathrm{ex}} P_{\tau}\re\left\{
  (1-\langle{\nu_1|\nu_2}\rangle)\langle{e_1|e_2}\rangle
\right\}I.
\end{equation}
Equation~(\ref{Paper::eq:I:5}) shows a sharp contrast between the roles
of the two types of decoherence.  The decoherence that is due to photon
emission and described in effect by $\ket{\nu_j}$ enhances the current
while the usual decoherence process (described by $\ket{e_j}$) due to
the coupling to the environment suppresses the current.


At this point, it will be interesting to address the question:
Does this spontaneous dc current violate the second law of
thermodynamics? Consider four atoms, two from each reservoir. One
will end up with (on average) three atoms in the reservoir 1 but
one in the reservoir 2, which corresponds to the decrease of the
entropy by $\log(3/2)$.
However, the increase in entropy induced by the decoherence is enough to
compensate this decrease and give a net increase in \emph{total} entropy.
To see this, note that the complete decoherence makes the
off-diagonal components of the density matrix zeros, which gives
rise to the increase of the entropy by $\log2$. Therefore the net
increase in total entropy is $(1/2)\log2-(1/4)\log(3/2)\approx\log
1.3$ per atom.

\begin{figure}
\center
\includegraphics*[width=65mm]{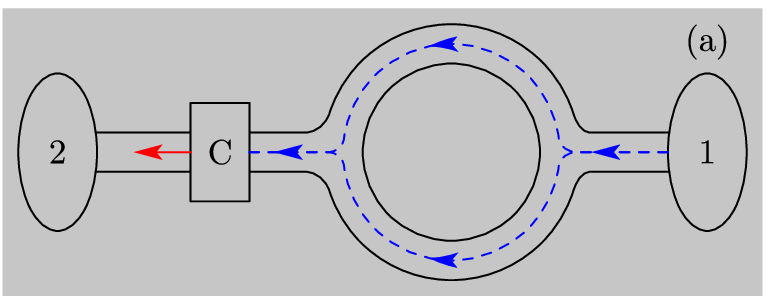}\\[2mm]
\includegraphics*[width=65mm]{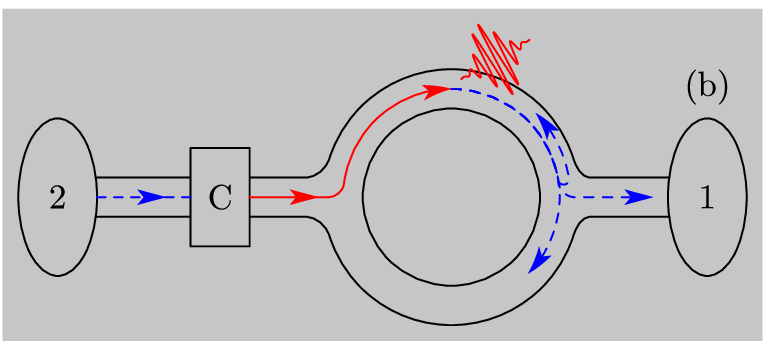}\\[2mm]
\caption{(color online) The scheme based on a ring geometry.
  (a) The coherent process: An atom the reservoir 1 experiences a
  coherent constructive interference and reaches the reservoir 2 with
  unit probability. The cavity does not affect the
  transmission of this atom.  (b) The incoherent process:
  An atom from the reservoir 2 is first excited at the
  cavity, and spontaneously emits a photon passing
  through the upper arm. The wave starting from the upper arm is scattered
  at the right junction.}
\label{Paper::fig:fig3}
\end{figure}

Now we turn to the second example,
i.e., the scheme based on the ring geometry; see
Fig.~\ref{Paper::fig:fig3}.
This scheme is interesting in the light that the coherent electron
transport through the ring has been extensively investigated in
condensed matter physics.
The basic principle is exactly the same as in the first scheme: The
coherent propagation [blue dashed line in
Fig.~\ref{Paper::fig:fig3}(a)] of the atoms from reservoir 1 enables
them to reach the reservoir 2 with unit probability.  The atoms from
the reservoir 2, on the other hand, are excited before entering the
interferometer, and experiences decoherence within the
interferometer; Fig.~\ref{Paper::fig:fig3}(b).  This incoherent
propagation reduce the probability for the atoms to reach the
reservoir 1.  Overall we have a net dc current from reservoir 1 to
2.
To estimate the amount of the current, we calculate the
transmission probability for incoherent process using the same
formalism as in the Michelson interferometer except that we
replace the scattering matrix in Eq.~(\ref{Paper::eq:S1}) for the
BS with that for the three-terminal junction \cite{Shapiro83a}:
\begin{equation}
\label{unitary_matrix}
J = \left( \begin{array}{ccc}
-(a+b) & \sqrt{\epsilon} & \sqrt{\epsilon} \\
\sqrt{\epsilon} & a & b \\
\sqrt{\epsilon} & b & a
\end{array} \right)
\end{equation}
with $a=(\sqrt{1-2\epsilon}-1)/2$ and $b=(\sqrt{1-2\epsilon}+1)/2$.
$\epsilon$ is the coupling parameter: $\epsilon=1/2$ gives no reflection
when the waves enter the ring from the reservoirs. For $\epsilon=0$,
the ring and the lead are completely decoupled. For simplicity, let us
assume that $\epsilon=1/2$.

Now suppose the atom emits a photon, say, in the lower arm.  The atom
(now in the ground-sate) starts coherent propagation.  Simple inspection
shows that the final transmission probability is 0.5, and there is no
reflection. Because of the conservation of the number of the particles,
the remaining one half should be trapped in the ring!

Let us consider this striking result more carefully. In Fig. 3(b)
the wave starting from the upper arm scatters on the right
junction. $\sqrt{1/2}$ of the wave amplitude (probability $1/2$)
escapes from the ring, $-1/2$ is reflected back to the upper arm,
and $1/2$ is transmitted to the lower. The phase accumulation
during the passage of either arm is ignored for the moment. The
two transmitted and the reflected waves are scattered again on the
left junction. Since these two waves are out of phase with the
same magnitude, they interferes destructively in the lead attached
to the reservoir 2. Repeating similar analysis shows that 50\% of
the wave is trapped in the ring when the spontaneous decay takes
place in the ring. 
In reality, due to various sources of decoherence the atom should
finally escape from the ring to the right or left reservoir with
the same probability. Roughly, 75\% of the incoming wave is
transmitted and 25\% reflected.  Compared with the coherent case,
25\% of the transmission probability decreases, which results in
the net dc current. The trapping probability depends on the
geometry of the ring such as $\epsilon$ and the length of the arm,
$L$ \cite{endnote:1}.

In conclusion, we have proposed a new mechanism to generate dc current
using a noble interplay between coherence and decoherence. Two specific
schemes of implementation have been presented based on the Michelson and
the ring interferometers.
A coherent superposition of states has more information (or equivalently
less entropy) than incoherent ones.  In some sense, this extra
information has been exploited to generate a dc current.
Thus it will be interesting to compare our work with another striking
proposal by Scully \textit{et al.}\cite{Scully03a}, a quantum heat
engine operating from a single heat bath prepared in a certain coherent
superposition and with a greater efficiency than a classical Carnot engine.
The idea presented here may hopefully shed light on deeper understanding
of the nature of decoherence and the subtle boundary between classical
and quantum physics.

\begin{acknowledgments}
We thank Hyun-Woo Lee and Markus B\"uttiker for helpful discussions.
SWK was supported by the KOSEF Grant (R01-2005-000-10678-0) and by the
KRF Grant (KRF-2004-005-C00044).
MSC was supported by the eSSC, the SKORE-A, and the SK Fund.
\end{acknowledgments}


\end{document}